\documentclass[aps,prb,superscriptaddress,twocolumn]{revtex4-2}

\usepackage{graphicx}
\usepackage{amsmath,amssymb}
\usepackage{bm}
\usepackage{hyperref} 

\begin{document}

\title{Electrostatic Lens of a Charged Vortex}

\author{Iogann Tolbatov}
\email{itolbatov@uniss.it}
\affiliation{Department of Chemical, Physical, Mathematical and Natural Sciences, University of Sassari, 07100 Sassari, Italy}

\author{Luca Salasnich}
\affiliation{Dipartimento di Fisica e Astronomia ``Galileo Galilei'' and Padua QTech Center, Universit\`a di Padova, Via Marzolo 8, 35131 Padova, Italy}
\affiliation{INFN Sezione di Padova, Via Marzolo 8, 35131 Padova, Italy}

\date{2 October 2026 --- Published in \emph{Condens. Matter} \textbf{11}(4), 35 (2026); DOI: 10.3390/condmat11040035}

\begin{abstract}
We study the local electrostatic response of a charged superfluid in the presence of a cylindrically symmetric vortex, allowing both the particle density and the thermodynamic compressibility to vary with the radial coordinate. Starting from the electrostatic charge response obtained from the phase-only Popov action, we introduce a local-density extension and derive the regular near-axis expansion of the scalar potential. The constant, quadratic, and quartic coefficients are obtained explicitly in terms of the local screening response and of the leading nonuniform contribution to the vortex density profile. For a three-dimensional unitary Fermi equation of state, the density dependence of the compressibility generates a definite thermodynamic contribution to the quartic coefficient. The resulting decomposition separates this equation-of-state contribution from the ordinary constant-screening and charge-source terms that occur at the same order. This provides a compact analytical connection between vortex-core thermodynamics and the local electrostatic profile of a charged superfluid.
\end{abstract}

\maketitle

\section{Introduction}

The problem of macroscopic electric field penetration in superconductors has a long history, tracing back to the phenomenological London equations \cite{London1935}. Thermodynamic and hydrodynamic approaches further established that quantized topological defects accumulate a local core charge due to chemical potential differences between the core and the bulk \cite{Blatter1996,Lipavsky2002}. A modern thermodynamic approach to this screening is provided by the phase-only Popov action, which describes the low-energy electrodynamics of a charged superfluid in terms of its global equation of state and the corresponding local compressibility \cite{Salasnich2024}. 

However, current electrostatic screening models derived from this framework typically assume a uniform bulk density. Quantized topological defects, such as vortices, naturally break this assumption because the local particle density is strongly suppressed near the core axis. In standard metals, the microscopic physics of the core is governed by Caroli--de Gennes--Matricon (CdGM) states \cite{CdGM1964}, which emerge from the Bogoliubov--de Gennes equations and trap normal-fluid quasiparticles near the axis \cite{deGennes1966,Shore1989}. These bound states maintain a finite, non-zero particle density $n_c > 0$ at the center, preventing non-analytic singular behavior in the thermodynamic screening. While this mechanism effectively maintains charge neutrality in standard high-density metals, the thermodynamic scaling changes drastically in strongly coupled superfluids near the BCS--BEC crossover \cite{Leggett1980,Nozieres1985}, where the density dependence of the equation of state becomes highly nonlinear.

In this work, we bridge this gap by introducing a local-density extension to the Popov action. In classical charged-particle optics, a cylindrically symmetric electrostatic field with a regular radial expansion $\Phi(r) = \Phi_0 + \Phi_2 r^2 + \Phi_4 r^4 + \dots$ acts as an electrostatic lens: the quadratic coefficient $\Phi_2$ governs paraxial focusing, while the quartic coefficient $\Phi_4$ determines the leading spherical aberration. By allowing the thermodynamic compressibility to vary radially with the density profile of a charged vortex, we derive the explicit analytical coefficients $\Phi_0$, $\Phi_2$, and $\Phi_4$ inside the core. This approach mathematically isolates the thermodynamic equation-of-state contribution to the aberration term from ordinary constant-screening effects, offering a fully controlled near-axis expansion.

\section{Local Electrostatic Equation}

For clarity, we briefly recall the corresponding uniform-density expressions established in Ref.~\cite{Salasnich2024}. In the absence of phase fluctuations, the zero-temperature internal energy density of a uniform superfluid Fermi gas is given by $\mathcal{E}_0(n) \propto n^{5/3}$ (specifically, $\mathcal{E}_0(n) = \frac{5}{3}(3\pi^2)^{2/3}n^{5/3}$ when neglecting residual interactions), which directly corresponds to the unperturbed uniform pressure equation of state $P(\mu) = \frac{2}{15\pi^2}\left(\frac{2m}{\hbar^2}\right)^{3/2}\mu^{5/2}$. Under the Popov prescription, the phase field $\theta(\mathbf{r},t)$ is introduced by shifting the chemical potential, yielding the effective action from which the electrodynamics are derived.

For a charged superfluid, the static charge density obtained to quadratic order in the scalar potential from this phase-only Popov action can be written as \cite{Salasnich2024}
\begin{equation}
\rho = q\bigl(n-\bar n_{\rm bg}\bigr)-q^2 P''(\mu)\Phi,
\label{eq:rho}
\end{equation}
where $q$ is the generic effective charge of the particles described by the superfluid theory (for example, $q = -e$ for fermionic charge carriers or $q = -2e$ if the theory is explicitly formulated in terms of Cooper-pair density, with $e > 0$ being the elementary charge), $n$ is their local particle-number density, $\bar n_{\rm bg}$ is the uniform number density of the neutralizing positive background, $\mu$ is the chemical potential, and $P''(\mu)=\partial^2 P/\partial\mu^2$. Equation~\eqref{eq:rho} is equivalent to Equation~(53) of Ref.~\cite{Salasnich2024}, using the thermodynamic identity
\begin{equation}
\frac{n}{c_s^2}=m P''(\mu),
\label{eq:cs}
\end{equation}
with $m$ the effective particle mass and $c_s$ the zero-temperature sound velocity.

For a homogeneous isotropic dielectric background with a spatially independent relative permittivity $\epsilon_r$, we use the effective-medium electrostatic equation
\begin{equation}
\nabla\cdot\mathbf E=\frac{\rho}{\epsilon},
\qquad
\epsilon\equiv\epsilon_0\epsilon_r,
\label{eq:gauss}
\end{equation}
where $\epsilon_0$ is the vacuum permittivity and $\mathbf E=-\nabla\Phi$. The replacement $\epsilon_0\to\epsilon_0\epsilon_r$ should be understood as a phenomenological effective-medium extension of the vacuum formulation of Ref.~\cite{Salasnich2024}.

Combining Equations~\eqref{eq:rho} and \eqref{eq:gauss} gives
\begin{equation}
\nabla^2\Phi(\mathbf r)-\kappa^2(\mathbf r)\Phi(\mathbf r)
=-\frac{q}{\epsilon}\left[n(\mathbf r)-\bar n_{\rm bg}\right],
\label{eq:master}
\end{equation}
where, within a local-density approximation,
\begin{equation}
\kappa^2(\mathbf r)
=\frac{q^2}{\epsilon}P''\!\left(\mu(\mathbf r)\right)
=\frac{q^2}{m\epsilon}\frac{n(\mathbf r)}{c_s^2(\mathbf r)}.
\label{eq:kappa}
\end{equation}
Thus, the equation of state enters the electrostatic problem through the local compressibility $P''(\mu)$.

For a straight line vortex, we assume translational invariance along the vortex axis and cylindrical symmetry in the transverse plane. Equation~\eqref{eq:master} then becomes
\begin{equation}
\frac{1}{r}\frac{d}{dr}\left(r\frac{d\Phi}{dr}\right)
-\kappa^2(r)\Phi(r)
=-\frac{q}{\epsilon}\left[n(r)-\bar n_{\rm bg}\right].
\label{eq:radial}
\end{equation}

\section{Equation of State and Near-Core Expansion}

We assume that the total particle density entering the effective equation admits a regular expansion about the vortex axis,
\begin{equation}
n(r)=n_c+n_2 r^2+\mathcal{O}(r^4),
\label{eq:nexp}
\end{equation}
where $n_c=n(0)$ and $n_2$ is the leading curvature of the density profile. Such a quadratic rise of the density near a vortex core is found, for example, in microscopic calculations across the BCS--BEC crossover \cite{Sensarma2006}. This $\mathcal{O}(r^2)$ density variation is characteristic of a singly quantized vortex (winding number $w=1$). For a multiply quantized giant vortex with winding number $w > 1$, the order parameter vanishes as $r^w$ near the axis, meaning the leading density variation scales instead as $r^{2w}$. In such higher-winding cases, $n_2 = 0$, which causes the compressibility gradient $C$ to vanish and pushes the thermodynamic equation-of-state signatures to higher orders (starting at $\Phi_{2w+2}$) in the electrostatic potential expansion. Returning to the $w=1$ case, we deliberately keep $n_c$ and $n_2$ as independent input parameters. If desired, one may write $n_2=\delta n/\xi^2$, with $\xi$ a characteristic core length and $\delta n$ a density scale, but no specific value of either quantity is required below.

For a general equation of state, Equation~\eqref{eq:kappa} has the regular expansion
\begin{equation}
\kappa^2(r)=\kappa_0^2+C r^2+\mathcal{O}(r^4),
\label{eq:kexp}
\end{equation}
with
\begin{equation}
\kappa_0^2=\kappa^2(n_c),
\qquad
C=\left.\frac{d\kappa^2}{dn}\right|_{n_c} n_2.
\label{eq:Cgeneral}
\end{equation}
The bare charge source appearing on the right-hand side of Equation~\eqref{eq:radial} is correspondingly
\begin{equation}
q\left[n(r)-\bar n_{\rm bg}\right]
=\rho_0+\rho_2 r^2+\mathcal{O}(r^4),
\label{eq:rhoexp}
\end{equation}
where
\begin{equation}
\rho_0=q\left(n_c-\bar n_{\rm bg}\right),
\qquad
\rho_2=q n_2.
\label{eq:rho02}
\end{equation}
For a neutral homogeneous bulk, $\bar n_{\rm bg}$ is fixed by the bulk particle density; it is not an independent charge source.

As an explicit example, consider the three-dimensional zero-temperature unitary Fermi-gas equation of state
\begin{equation}
P(\mu)=K\mu^{5/2},
\label{eq:unitaryP}
\end{equation}
with
\begin{equation}
K=\frac{2}{15\pi^2}
\left(\frac{2m}{\hbar^2}\right)^{3/2}\xi_B^{-3/2},
\label{eq:K}
\end{equation}
where $\xi_B$ is the Bertsch parameter \cite{Zwerger2012,Ku2012}. From $n=P'(\mu)$, one obtains
\begin{equation}
\mu=\left(\frac{2n}{5K}\right)^{2/3},
\qquad
P''(\mu)=\gamma n^{1/3},
\label{eq:unitaryderivatives}
\end{equation}
where
\begin{equation}
\gamma=\frac{15}{4}K\left(\frac{2}{5K}\right)^{1/3}.
\end{equation}
Therefore,
\begin{equation}
\kappa^2(n)=A n^{1/3},
\qquad
A=\frac{q^2\gamma}{\epsilon},
\label{eq:unitarykappa}
\end{equation}
and Equation~\eqref{eq:Cgeneral} gives
\begin{equation}
C=\frac{\kappa_0^2}{3n_c}n_2.
\label{eq:Cunitary}
\end{equation}
If $n_2=\delta n/\xi^2$, this becomes $C=\kappa_0^2\delta n/(3n_c\xi^2)$.

The same reasoning can be stated for a power-law equation of state $P(\mu)=K\mu^\alpha$ with $\alpha>1$. Defining
\begin{equation}
\beta=\frac{\alpha-2}{\alpha-1},
\end{equation}
one finds exactly
\begin{equation}
\kappa^2(n)=B_\alpha n^\beta,
\end{equation}
with
\begin{equation}
B_\alpha=\frac{q^2}{\epsilon}\alpha(\alpha-1)K
\left(\alpha K\right)^{-\beta},
\end{equation}
and hence
\begin{equation}
C=\beta\,\frac{\kappa_0^2}{n_c}n_2.
\label{eq:Cpower}
\end{equation}
For $\alpha=2$, $P''(\mu)$ is constant and the equation-of-state contribution $C$ vanishes. This does not imply that the full quartic coefficient of the potential vanishes, because other contributions remain at the same order.

\section{Near-Axis Electrostatic Potential}

Substituting Equations~\eqref{eq:kexp} and \eqref{eq:rhoexp} into Equation~\eqref{eq:radial} gives, through order $r^2$,
\begin{equation}
\frac{1}{r}\frac{d}{dr}\left(r\frac{d\Phi}{dr}\right)
-\left(\kappa_0^2+C r^2\right)\Phi
=-\frac{1}{\epsilon}\left(\rho_0+\rho_2 r^2\right).
\label{eq:inner}
\end{equation}
Regularity at the vortex axis requires $d\Phi/dr=0$ at $r=0$, so that
\begin{equation}
\Phi(r)=\Phi_0+\Phi_2r^2+\Phi_4r^4+\mathcal{O}(r^6).
\label{eq:Phiexp}
\end{equation}
The constant $\Phi_0=\Phi(0)$ is not determined by the local expansion alone. Its value must be fixed by the global electrostatic boundary conditions and by matching the inner solution to the solution outside the core.

\begin{figure}[htbp]
\centering
\includegraphics[width=0.48\textwidth]{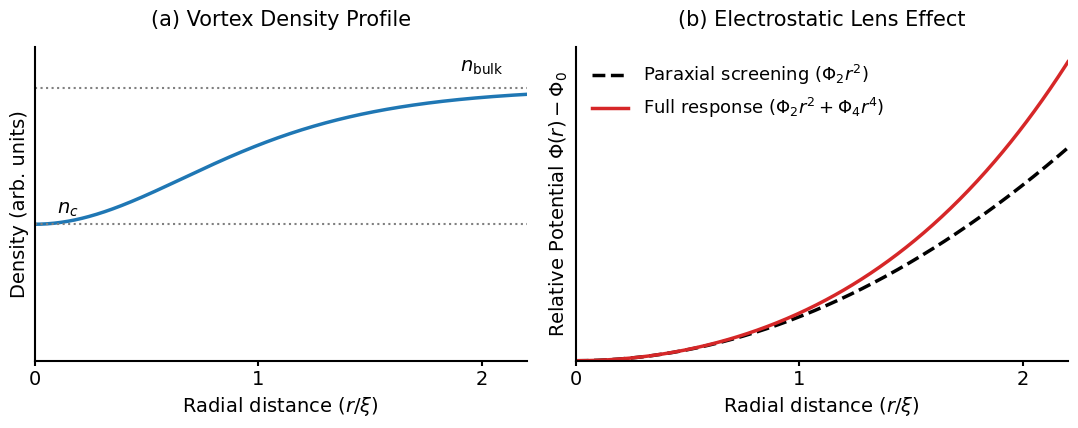}
\caption{Conceptual schematic of the local electrostatic lens effect in a charged vortex core. (a) The local particle density $n(r)$ exhibits a characteristic quadratic depletion near the axis, dropping from the bulk density $n_{\rm bulk}$ to a finite central core density $n_c$. (b) The resulting relative electrostatic potential $\Phi(r) - \Phi_0$. The purely quadratic screening term ($\Phi_2 r^2$, dashed line) acts as a standard paraxial lens, while the full local response (solid line) includes the quartic aberration term ($\Phi_4 r^4$), which directly encodes the thermodynamic equation-of-state contribution. Plotting the relative potential isolates the local near-axis physics, remaining independent of the global boundary conditions required to fix the absolute central potential $\Phi_0$.}
\label{fig:schematic}
\end{figure}

Applying the radial Laplacian to Equation~\eqref{eq:Phiexp} gives $4\Phi_2+16\Phi_4r^2+\mathcal{O}(r^4)$. At order $r^0$, we obtain
\begin{equation}
4\Phi_2-\kappa_0^2\Phi_0=-\frac{\rho_0}{\epsilon},
\end{equation}
and therefore
\begin{equation}
\Phi_2=
\frac{\kappa_0^2\Phi_0}{4}
-\frac{\rho_0}{4\epsilon}.
\label{eq:Phi2}
\end{equation}
At order $r^2$,
\begin{equation}
16\Phi_4-\kappa_0^2\Phi_2-C\Phi_0
=-\frac{\rho_2}{\epsilon},
\end{equation}
which gives
\begin{equation}
\Phi_4=
\frac{\kappa_0^4\Phi_0}{64}
-\frac{\kappa_0^2\rho_0}{64\epsilon}
+\frac{C\Phi_0}{16}
-\frac{\rho_2}{16\epsilon}.
\label{eq:Phi4}
\end{equation}
Equations~\eqref{eq:Phiexp}, \eqref{eq:Phi2} and \eqref{eq:Phi4} are the main result of this work, conceptually illustrated in Fig.~\ref{fig:schematic}.

The terms in Equation~\eqref{eq:Phi4} have a simple interpretation. The first two are already present for a constant screening coefficient and correspond to the local series of the ordinary cylindrical screened solution. The third term, $C\Phi_0/16$, is generated by the spatial dependence of the compressibility, while the fourth term is produced directly by the $r^2$ part of the bare charge source. Thus the presence of an $r^4$ term in $\Phi(r)$ is generic; the equation of state is encoded in a particular part of its coefficient.

For the unitary equation of state, Equation~\eqref{eq:Cunitary} gives an equation-of-state contribution $\Phi_4^{({\rm EOS})} = \kappa_0^2\Phi_0 n_2/(48n_c)$. If the bare central charge density $\rho_0$ is retained exactly, the size of this thermodynamic contribution relative to the leading parabolic term is given by:
\begin{equation}
\frac{\Phi_4^{({\rm EOS})}r^4}{\Phi_2r^2}
= \frac{n_2 r^2}{12n_c} \left(1 - \frac{\rho_0}{\epsilon\kappa_0^2\Phi_0}\right)^{-1}.
\label{eq:localratio}
\end{equation}
When the local charge depletion is significant ($n_c \ll \bar{n}_{\rm bg}$), $\rho_0$ is not small, and the relative magnitude of the thermodynamic contribution remains explicitly coupled to the central potential $\Phi_0$, which must be fixed by global boundary conditions. However, in the limit where the bare core charge is negligible compared to the electrostatic screening term ($|\rho_0/\epsilon| \ll \kappa_0^2|\Phi_0|$), the central potential $\Phi_0$ explicitly cancels out. In this decoupled regime, if one writes $n_2=\delta n/\xi^2$, the ratio yields $(\delta n/n_c)(r/\xi)^2/12$. This limit illustrates analytically that the quartic contribution behaves as a controlled correction near the axis rather than a replacement of the quadratic term (see Fig.~\ref{fig:schematic}).

The corresponding radial electric field and total charge density are
\begin{align}
E_r(r)&=-2\Phi_2r-4\Phi_4r^3+\mathcal{O}(r^5),\\
\rho_{\rm tot}(r)&=\epsilon\,\nabla\cdot\mathbf E
=-4\epsilon\Phi_2-16\epsilon\Phi_4r^2+\mathcal{O}(r^4).
\end{align}
Hence, the quartic term of the potential appears as a cubic correction in the electric field and as a quadratic correction in the total charge density.

\section{Discussion and Physical Implications}

The local expansion derived above provides a direct way of connecting vortex-core structure to electrostatic screening. Once the first coefficients of the density profile are known, either from a microscopic vortex calculation or from an independent effective description, the electrostatic coefficients follow algebraically. In this sense, the result is complementary to microscopic Bogoliubov--de Gennes \cite{Kopnin1996,Ohuchi2022} or density-functional approaches: such calculations can supply the local density profile, while the present framework isolates how the thermodynamic compressibility enters the electrostatic response.

The local-density construction is particularly useful because the separation of the quartic coefficient into constant-screening, source, and equation-of-state contributions does not depend on a specific global vortex ansatz. A more detailed core profile can change the numerical values of the local density coefficients, but the same order-by-order procedure can be applied once those coefficients are specified. The analytical structure therefore provides a convenient benchmark for more microscopic calculations of charged vortices. 

We emphasize that this framework strictly relies on the validity of the local-density approximation, which requires careful consideration of the underlying length scales. For the dielectric permittivity $\epsilon(r)$, the approximation is robust: background polarizability is driven by atomic-scale interband transitions, which are physically decoupled from the macroscopic coherence length $\xi$ over which the vortex varies. For the thermodynamic compressibility, the validity assumes $\xi$ is large compared to the interparticle spacing $k_F^{-1}$, making $(k_F \xi)^{-1}$ the controlling small parameter. If one introduces leading microscopic gradient corrections to the equation of state (e.g., an energy density correction proportional to $|\nabla n|^2$), the resulting shift in the local chemical potential introduces a term proportional to $\nabla^2 n$. Given the near-axis density profile $n(r) = n_c + n_2 r^2 + \mathcal{O}(r^4)$, this Laplacian generates an $\mathcal{O}(r^2)$ correction to the local compressibility. Because this gradient modification to the compressibility scales precisely as $r^2$, it enters the electrostatic equation exactly parallel to the local-density parameter $C r^2$. Consequently, gradient corrections compete strictly at the same $\mathcal{O}(r^4)$ order in the scalar potential. While these gradient effects are heavily suppressed by $(k_F \xi)^{-2}$ in the BCS regime, they become quantitatively comparable to the pure local-density contribution in strongly coupled superfluids where $\xi \sim k_F^{-1}$.

The dielectric response can likewise be adapted to the physical platform of interest. In the present treatment, it is represented by a homogeneous scalar permittivity, which is sufficient to display explicitly how electrostatic screening competes with the thermodynamic response. For a material-specific calculation, the same construction can be combined with an experimentally determined or microscopically calculated dielectric response. This is especially relevant in low-carrier-density superconductors \cite{Coldea2018,Shibauchi2020}, where both compressibility and screening can differ substantially from those of conventional metals.

The unitary Fermi gas offers a particularly transparent analytical example because its nonlinear equation of state produces a density-dependent screening coefficient already at the leading nonuniform order. More generally, any equation of state with a nonconstant compressibility generates an analogous thermodynamic contribution to the quartic electrostatic coefficient. The vanishing of that contribution for a quadratic equation of state provides a useful contrast and shows explicitly which part of the near-core potential is controlled by thermodynamics rather than by geometry alone.

\section{Conclusions}

We have obtained an analytical near-axis description of the electrostatic potential generated by a cylindrically symmetric charged vortex within a local-density extension of the phase-only electrostatic response. The central result is the explicit determination of the constant, quadratic, and quartic coefficients of the potential. The quartic coefficient naturally separates into contributions from ordinary screening, from the spatially varying charge source, and from the density dependence of the thermodynamic compressibility. This decomposition makes it possible to identify directly the part of the electrostatic profile that carries information about the equation of state.

The three-dimensional unitary Fermi gas provides a simple strong-coupling example in which the compressibility produces a nonzero thermodynamic quartic contribution. The same construction applies more broadly to charged superfluids with a nonlinear equation of state and therefore offers a compact analytical benchmark for comparing different interaction regimes and microscopic vortex models.

To demonstrate the physical scaling of these analytical results, bulk FeSe serves as an ideal platform. FeSe is a benchmark compensated semimetal in the BCS--BEC crossover~\cite{Hsu2008,Kasahara2016} with an extremely low bulk carrier density, typically $n_{\rm bulk} \approx 10^{20}$ to $10^{21}$~cm$^{-3}$ \cite{Hanzawa2016}. In our framework, the local expansion is parameterized by the core density at the axis, $n_c \equiv n(0)$; for a physical estimate, we assume both $n_c$ and the spatial density variation $\delta n$ are on the order of $n_{\rm bulk}$. The numerical evaluation assumes elementary charge carriers ($q = -e$), an effective mass $m \approx m_e$ appropriate for the shallow pockets, and a homogeneous neutralizing background $\bar{n}_{\rm bg} = n_{\rm bulk}$. Furthermore, FeSe exhibits an unusually large static background dielectric constant of $\epsilon_r \approx 65$ \cite{Chowdhury2016}. Because the superconducting transition only involves states very close to the Fermi level, this lattice polarizability remains highly applicable inside the superconducting phase. Because of this combination of high background permittivity and low carrier concentration, the Thomas--Fermi screening length is significantly enhanced compared to conventional metals, reaching $\lambda_E \approx 0.9\text{--}1.4$~nm. Since this screening length remains strictly smaller than the in-plane healing length ($\lambda_E < \xi$, with $\xi \approx 5.5$~nm \cite{Kasahara2014}), the background density varies slowly over the characteristic scale of the electrostatic response. Consequently, the local-density approximation utilized for our vortex core expansion is physically well controlled. We note that while our explicit analytical example of the unitary equation of state strictly assumes zero temperature, the experimental realization of FeSe necessarily occurs at finite temperature; the subsequent evaluation should therefore be understood as a qualitative order-of-magnitude estimate of the thermodynamic scaling.

Using this established scale separation, and provided the bare core charge remains negligible compared to the screening term, we can apply our newly derived local ratio to estimate the size of the thermodynamic effect near the core. The relative magnitude of the thermodynamic quartic contribution compared to the leading parabolic screening term is given by:
\begin{equation}
\frac{\Phi_4^{({\rm EOS})} r^4}{\Phi_2 r^2} \simeq \frac{1}{12} \frac{\delta n}{n_c} \left(\frac{r}{\xi}\right)^2
\end{equation}
If we evaluate this at the edge of the vortex core ($r = \xi$), the ratio simplifies to $\frac{1}{12} \frac{\delta n}{n_c}$. Even if the dielectric background of FeSe ($\epsilon_r = 65$) allows for a substantial local density depletion (e.g., $\delta n \approx n_c$), the thermodynamic quartic contribution represents a maximum perturbation of roughly $8\%$ relative to the standard parabolic screening term. This perfectly satisfies the requirement of analytical control: it proves that the $\mathcal{O}(r^4)$ effect is a mathematically consistent and systematic correction near the axis.

\begin{acknowledgments}
L.S. is partially supported by the Project ``Frontiere Quantistiche'' (Dipartimenti di Eccellenza) of the Italian Ministry of University and Research and by ``Iniziativa Specifica Quantum'' of INFN.
\end{acknowledgments}

\end{document}